\documentclass[
    ,final            
  ]
  {aipproc}

\layoutstyle{8x11single}

\begin{document}

\title{Classical Electrodynamics from the Motion of a Relativistic Fluid}



\author{Sabbir A. Rahman}{
  address={Theoretical Physics, Blackett Laboratory, Imperial College of Science, Technology \& Medicine, Prince
Consort Road, London SW7 2BZ, U.K.} }

%

\begin{abstract}
We show that there exists a choice of gauge in which the
electromagnetic 4-potential may be written as the difference of
two 4-velocity vector fields describing the motion of a
two-component space-filling relativistic fluid. Maxwell's
equations are satisfied immediately, while the Lorentz force
equation follows from the interactions of sources and sinks. The
usual electromagnetic quantities then admit new interpretations as
functions of the local 4-velocities. Electromagnetic waves are
found to be described by oscillations of the underlying medium
which can therefore be identified with the `luminiferous aether'.
The formulation of electrodynamics in terms of 4-velocities is
more general than that of the standard 4-potential in that it also
allows for a classical description of a large class of vacuum
energy configurations. Treated as a self-gravitating fluid, the
model can be explicitly identified with Nelson's stochastic
formulation of quantum mechanics, making it a promising candidate
as the classical field theory unifying gravitation,
electromagnetism and quantum theory which Einstein had sought.
\end{abstract}

\maketitle


\section{Introduction}

Motivated by the fact that macroscopic waves tend to propagate in
a medium of some kind, many physicists have in the past attempted
to find a description of electrodynamics in which electromagnetic
waves may also be described by the motion of an underlying medium,
commonly referred to as the `aether'\cite{Whittaker53}. Despite
significant efforts, no such description was found, and the
negative results of the Michelson-Morley
experiment\cite{Swenson72} put the final nail in the coffin of the
aether concept as it was then understood. The failure of this
program eventually led to the introduction of the concept of a
`field'\cite{Maxwell73} requiring no underlying medium, and on
quantisation of these fields, to `quantum field theory'.

We demonstrate here that Maxwell's equations can in fact be
derived from the motion of an underlying medium if we assume from
the outset that the underlying spacetime is Lorentzian rather than
Galilean. While our formulation differs from earlier conceptions
of the aether, it does show that an alternative description of
electrodynamics did (and does) exist which did not require the
introduction of the field concept. The medium can be interpreted
as a two-component relativistic fluid with sources and sinks
playing the role of charges. This fluid dynamical model allows an
explicit connection to be made with Nelson's stochastic
formulation of quantum mechanics, suggesting that the foundations
of quantum electrodynamics may lie purely in classical gravity as
Einstein had long believed.

We assume a metric with signature $(+, -, -, -)$, and follow the
conventions of Jackson\cite{JDJ98} throughout.

\section{The Relativistic Continuum}

We first establish the reference system and coordinates that will
be used to describe the motion of the continuum. We then define
the `continuum gauge', showing how Maxwell's equations equations
appear. We also show that the Lorentz force equation translates
into a third order partial differential equation constraining the
continuum's motion.

\subsection{Coordinates and Reference Frames}

Consider an arbitrary relativistic inertial frame with
4-coordinates $x^\mu = (ct, x, y, z)$, so that the spacetime
partial derivatives are given by $\partial^\mu = ({1\over c}
{\partial\over\partial t}, - {\bf\nabla})$ and $\partial_\mu =
({1\over c} {\partial\over\partial t}, {\bf\nabla})$ respectively.
Suppose that there exists a continuous medium in relative motion
to this frame which spans the entire spacetime. To describe
positive and negative charge configurations we will see that this
continuum must consist of two superimposed but mutually
independent components which we shall refer to as the `positive
continuum' and the `negative continuum' respectively.

Let $\tau$ be the proper time in the inertial frame, and consider
the instantaneous motion at proper time $\tau$ of either component
of the continuum at any 3-position ${\bf r}=(x, y, z)$. Then the
3-velocity of that component at ${\bf r}$ as measured by the
inertial frame is,

\begin{equation}
{\bf v} = {d{\bf r}\over dt}\,,\label{eqn:velocity}
\end{equation}

\noindent where $t$ is the time as measured by a clock moving with
the continuum component. We can therefore define the interval,

\begin{equation}
ds^2 \equiv c^2 d\tau^2 = c^2 dt^2 - dx^2 - dy^2 -
dz^2\,.\label{eqn:interval}
\end{equation}

\noindent Similarly, we can define a 4-velocity vector field
describing the motion of the continuum component as,

\begin{equation}
u^\mu = {dx^\mu\over d\tau} = \left(c {dt\over d\tau}, {d{\bf
r}\over d\tau}\right) = (c \gamma,
\gamma\bf{v})\,,\label{eqn:fourvel}
\end{equation}

\noindent where $\gamma = (1 - v^2/c^2)^{-1/2}$ is the Lorentz
factor at each point. This 4-velocity clearly satisfies,

\begin{equation}
u^\mu u_\mu = c^2, \qquad u^{\mu, \nu} u_\mu = 0\,,\label{eqn:sol}
\end{equation}

\noindent where partial derivatives ${\partial^\nu u^\mu}$ are
written as $u^{\mu, \nu}$ for convenience.

The above definitions and identities hold for both the positive
and negative continua independently, and henceforth we distinguish
the two sets of variables by a `+' or `-' subscript respectively
as necessary.

\subsection{The Continuum Gauge}

The key to what follows is to split the electromagnetic potential
4-vector $A^\mu$ into the sum of two components $A_+^\mu$ and
$A_-^\mu$, which we identify (up to a dimensional constant $k$)
with the two continuum 4-velocities $u_+^\mu$ and $u_-^\mu$,

\begin{equation}
A^\mu = A_+^\mu + A_-^\mu\,,\qquad\hbox{where}\quad A_+^\mu = k
u_+^\mu = (\phi_+, {\bf A}_+)\,,\quad A_-^\mu = -ku_-^\mu =
(\phi_-, {\bf A}_-)\,.\label{eqn:fourpot}
\end{equation}

\noindent 
The condition \eqref{eqn:sol} implies the following covariant
constraint for both $A_+^\mu$ and $A_-^\mu$,

\begin{equation}
A_+^\mu A_{+ \mu} = A_-^\mu A_{- \mu} = k^2c^2\,.\label{eqn:gauge}
\end{equation}

We will refer to conditions \eqref{eqn:fourpot} and
\eqref{eqn:gauge} as the `continuum gauge'. This is a non-standard
choice of gauge, and we will demonstrate its consistency in the next
section. The antisymmetric field-strength tensor can now be defined
as,

\begin{equation}
F^{\mu\nu} = A^{\nu, \mu} - A^{\mu, \nu}\ \sim ({\bf E}, {\bf
B})\,.\label{eqn:stresstensor}
\end{equation}

\noindent Other standard properties now follow in the usual way.
From the definition \eqref{eqn:stresstensor}, $F^{\mu\nu}$
satisfies the Jacobi identity,

\begin{equation}
F^{\mu\nu, \lambda} + F^{\nu\lambda, \mu} + F^{\lambda\mu, \nu} =
0\,,\label{eqn:jacobi}
\end{equation}

\noindent and this is just the covariant form of the homogeneous
Maxwell's equations. One can \emph{define} the 4-current as the
4-divergence of the field-strength tensor,

\begin{equation}
J^\mu = {c\over4\pi} F^{\mu\nu}_{,\nu}\ = (c \rho, {\bf
j})\,,\label{eqn:current}
\end{equation}

\noindent and this is the covariant form of the inhomogeneous
Maxwell's equations. Charge conservation is guaranteed by the
antisymmetry of $F^{\mu\nu}$. The covariant Lorentz force equation
takes the following form,

\begin{equation}
{dV^\mu\over d\tau} = {Q\over M c} F^{\mu\nu}
V_\nu\,,\label{eqn:lorentz}
\end{equation}

\noindent where $Q$, $M$ and $V^\mu = (c \gamma_V, \gamma_V {\bf
V})$ are the charge, mass and 4-velocity vector of the observed
particles. This cannot be derived directly from the definition of
the 4-potential, and must be considered for now as an auxiliary
constraint.

\noindent The charge 4-velocity $V^\mu$ and scalar charge $Q$ are
related to the 4-current density $J^\mu$ through the following
equation,

\begin{equation}
J^\mu = Q V^\mu\,,\qquad\hbox{where}\quad V^\mu V_\mu =
c^2\,,\quad V^0\geq c\,.\label{eqn:charge}
\end{equation}

\noindent The constraint on $V^\mu$ allows us to separate the
4-current uniquely into the charge and its 4-velocity. Indeed we
have,

\begin{equation}
Q = \hbox{sgn}(J^0) \cdot \left({1\over c^2} J^\mu
J_\mu\right)^{1/2}\,,\label{eqn:zerocomp}
\end{equation}

\noindent where the sign of the 0-component of the 4-current
appears to ensure that the 0-component $V^0$ of the charge
4-velocity is positive. Since the sign of $J^0$ cannot be flipped
by a Lorentz transformation, each 4-velocity vector field can only
account for either positive or negative charge configurations, and
hence the need for two separate continuum components, each with
opposite contributions to the 4-potential. The gauge based upon a
single 4-vector field was precisely that introduced by Dirac in
his classical model of the electron\cite{Dirac51a}, and it is
noteworthy that he was also led to speculate that this 4-velocity
field described the motion of a real, physical,
`aether'\cite{Dirac51b}. The form of the charge 4-velocity in
terms of the continuum 4-velocity now follows directly from
(\ref{eqn:charge}).

Besides the mass $M$ which is determined by initial conditions, each
of the terms in (\ref{eqn:lorentz}) may be written in terms of the
$4$-velocities $u_+^\mu$ and $u_-^\mu$. From the definitions of
$F^{\mu\nu}$, $J^\mu$, $Q$ and $V^\mu$, we find that the Lorentz
force equation (\ref{eqn:lorentz}) translates into a complicated
third order partial differential equation constraining the
$4$-velocities. The conservation of mass follows from the continuity
equation for mass density,

\begin{equation}
(M V^\mu)_{,\mu} = 0\,,\label{eqn:masscons}
\end{equation}

\noindent which is ensured if the flow of mass density follows the
flow of charge density. We will see later that the Lorentz force
equation follows from the fluid dynamical interactions between
sources and sinks, and this will complete our picture of classical
electrodynamics in this gauge.

\section{The Consistency of the Continuum Gauge}

We have identified the components $A_+^\mu$ and $A_-^\mu$ of the
4-potential with the 4-velocities $u_+^\mu$ and $u_-^\mu$ of the
continuum satisfying the conditions (\ref{eqn:fourpot}) and
(\ref{eqn:gauge}), and have referred to this gauge choice as the
`continuum gauge'. It is not obvious that this gauge choice can be
applied consistently to all electromagnetic field configurations,
so we demonstrate its consistency here, and give explicit
solutions given for the point charge and the plane electromagnetic
wave.

\subsection{Proof of Consistency}

In order to prove consistency, it is necessary to find a
decomposition of the 4-potential as the difference of two
4-velocity fields satisfying equations (\ref{eqn:fourpot}) and
(\ref{eqn:gauge}) simultaneously. Using the notation of
(\ref{eqn:fourvel}), we therefore need to find, given any
4-potential $A^\mu=(\phi,{\bf A})$ defined up to a gauge
transformation $A^\mu\rightarrow A^\mu+\partial^\mu\psi$, two
3-velocity fields ${\bf v}_+$ and ${\bf v}_-$ satisfying the
following conditions,

\begin{equation}
{\phi\over kc} = \gamma_+-\gamma_-\,,\qquad{{\bf A}\over k} =
\gamma_+{\bf v}_+-\gamma_-{\bf v}_-\,.\label{eqn:gammadiff}
\end{equation}

The second of these equations is a simple geometrical vector
identity, and it is clear that any solution set for $(\gamma_+{\bf
v}_+, \gamma_-{\bf v}_-)$ will form a surface of revolution about
the axis defined by ${\bf A}$. To find the solution surface
explicitly for a given $(\phi,{\bf A})$, it is convenient to take
the origin to lie at ${\bf A}/2k$, and to use polar coordinates
$(r,\theta)$ in any plane containing ${\bf A}$, where
$r\in[0,\infty]$ is the radial distance from the origin and
$\theta\in[0,\pi]$ is the angle made with respect to the direction
of ${\bf A}$. Note the following simple chain of identities,

\begin{equation}
\gamma = {1\over\sqrt{1-{v^2\over c^2}}} \Rightarrow v =
c\,\sqrt{1-{1\over\gamma^2}} \Rightarrow \gamma v =
c\sqrt{\gamma^2-1} \Rightarrow \gamma = \sqrt{1+\left({\gamma
v\over c}\right)^2}\,,\label{eqn:gammidents}
\end{equation}
so that from (\ref{eqn:gammadiff}) we have,
\begin{equation}
{\phi\over kc} = \sqrt{1+\left({\gamma_+v_+\over c}\right)^2} -
\sqrt{1+\left({\gamma_-v_-\over
c}\right)^2}\,.\label{eqn:gammadiff2}
\end{equation}
Applying standard trigonometric identities to our geometrical
picture, we obtain,
\begin{equation}
(\gamma_+v_+)^2 = r^2+A^2/4k^2+{Ar\over k}
\cos\theta\,,\qquad(\gamma_-v_-)^2 = r^2+A^2/4k^2-{Ar\over k}
\cos\theta\,,\label{eqn:gv}
\end{equation}
so that the set of solutions on the plane in question is
determined by the condition,
\begin{equation}
\phi = \sqrt{A^2/4+Akr\cos\theta+k^2(r^2+c^2)} -
\sqrt{A^2/4-Akr\cos\theta+k^2(r^2+c^2)}\,.\label{eqn:phisol}
\end{equation}
Note that given any solution for $(\phi,{\bf A})$, a solution for
$(-\phi,{\bf A})$ is obtained by letting
$\theta\rightarrow\pi-\theta$. Note also (i) that $\phi=0$
whenever $\theta=\pi/2$ including when $r=0$, (ii) that for a
given value of $r$ the magnitude of $\phi$ is maximum when
$\theta=0$, (iii) that for $\theta=0$, $\phi$ is a monotonically
increasing function of $r$, and (iv) that $\phi\rightarrow
A\cos\theta$ as $r\rightarrow\infty$.

In conclusion, for a given value of $A=|{\bf A}|$, equations
(\ref{eqn:gammadiff}) will have solutions whenever $|\phi| \leq
A$. In the special case $\phi=0$ the solution surface for
$\gamma_+{\bf v}_+$ is just the plane perpendicular to ${\bf A}$
passing through the point ${\bf A}/2k$, throughout which $|{\bf
v}_+|=|{\bf v}_-|$, and $|\gamma_+{\bf v}_+|\geq A/2k$. For other
values of $|\phi| \leq A$ the solutions form a paraboloid-like
surface of revolution about the ${\bf A}$ axis. The sign of $\phi$
determines which side of the $\theta=\pm\pi/2$ plane the solution
surface lies.

It is always possible to choose the function $\psi$ defining the
choice of gauge in such a way that $\phi=0$
everywhere\cite{Landau75}. Since solutions to
(\ref{eqn:gammadiff}) always exist in this case, this proves that
the continuum gauge is indeed a consistent one.

It is important to note that there is actually a significant
additional degree of freedom inherent in the way the decomposition
of $A^\mu$ is made into 4-velocity fields, which goes beyond the
standard gauge freedom. First of all, for each electromagnetic
configuration there will be a continuum of gauge choices for which
a continuum gauge solution set exists. Secondly, for any
particular choice of gauge for which a solution does exist, there
will in general be an entire two-parameter surface of possible
solutions for ${\bf v}_+$ and ${\bf v}_-$ at each point in space.
We will show later that these velocity vector fields correspond to
the motion of massive discrete particles, so that this freedom may
have a real physical significance as a possible classical source
of dark matter.

\subsection{The Point Charge}

Let us now find the vacuum configuration which describes a
positive charge $q$ positioned at the origin. The corresponding
electromagnetic fields are given by,

\begin{equation}
{\bf E} = {q{\bf{\hat r}}\over r^2} = -\nabla\left({q\over
r}\right)\,,\qquad {\bf B} = 0\,.\label{eqn:eb}
\end{equation}

\noindent We seek a 4-potential of the following form which only
has contributions from the motion of the positive continuum,

\begin{equation}
A_+^\mu = (\phi_+, {\bf A}_+) = (kc \gamma, k\gamma {\bf
v})\,,\qquad A_-^\mu = (\phi_-, {\bf A}_-) = (-kc, {\bf
0})\,,\label{eqn:seekpot}
\end{equation}

\noindent where the velocity vector field ${\bf v}$ is to be
found. The corresponding electromagnetic fields ${\bf E}$ and
${\bf B}$ are given by,

\begin{equation}
{\bf E} = -\nabla\phi_+ - {1\over c} {\partial {\bf
A}_+\over\partial t} = - \nabla (kc \gamma) - {1\over c}
{\partial\over\partial t}(k\gamma {\bf v})\,,\qquad{\bf B} =
\nabla\times{\bf A}_+ = \nabla\times(k\gamma {\bf
v})\,.\label{eqn:ebfield}
\end{equation}

For any electrostatic configuration with stationary charges we
have ${\bf B} = \nabla\times(k\gamma{\bf v}) = 0$, so there must
exist a scalar field $\psi$ such that $k\gamma{\bf v}=\nabla\psi$.
After some algebraic manipulation this can be seen to imply that,

\begin{equation}
{v\over c} = {\nabla\psi\over\sqrt{k^2c^2 +
(\nabla\psi)^2}}\leq1\,,\qquad\gamma(v) = \left(1 +
{(\nabla\psi)^2\over k^2c^2}\right)^{1/2}\,,\label{eqn:vc}
\end{equation}
so that in terms of $\psi$, the ${\bf E}$ field is given by,

\begin{equation}
{\bf E} = - \nabla\left(\left(k^2c^2 +
(\nabla\psi)^2\right)^{1/2}\right) - {1\over c}
{\partial\over\partial t}
\left(\nabla\psi\right)\,.\label{eqn:epsi}
\end{equation}

Because of the rotational and time invariance of the problem, we
need only look for solutions of the form $\psi = \psi(r)$, so that
$\nabla\psi = \partial\psi/\partial r$ and the second term of
(\ref{eqn:epsi}) vanishes. Comparing with (\ref{eqn:eb}), it is
clear that $\psi$ must satisfy,

\begin{equation}
\left(k^2c^2 + \left({\partial\psi\over\partial
r}\right)^2\right)^{1/2} = {q\over r} +
\alpha\,,\label{eqn:psiconst}
\end{equation}

\noindent where $\alpha$ is a constant of integration. Since the
charge is positive and the velocity of the continuum should vanish
at infinity, we require $\alpha = kc$ for a real solution to
exist. From (\ref{eqn:psiconst}), the resulting differential
equation for $\psi$ is as follows,

\begin{equation}
{\partial\psi\over\partial r} = \pm\left(\left({q\over r} +
kc\right)^2 - k^2c^2\right)^{1/2}\,,\label{eqn:dpsidr}
\end{equation}

\noindent where either the positive or negative square root may be
chosen, as the 4-potential depends only on the magnitude of the
velocity and not its direction. There is therefore insufficient
information to specify whether the positive charge acts as a
source or a sink (or both). The solution for the velocity field
and the corresponding Lorentz factor is therefore,

\begin{equation}
{v\over c} = \pm \left(1 - \left(1 + {q\over krc}
\right)^{-2}\right)^{1/2}\,,\quad\quad\gamma=1+{q\over
krc}\,.\label{eqn:vcfinal}
\end{equation}
Note that $q/krc$ becomes singular at the origin, implying that
the continuum velocity in (\ref{eqn:vcfinal}) becomes equal to $c$
there.

The above confirms that the electromagnetic fields outside a
positive point charge can indeed be described by the motion of the
positive continuum, and that the corresponding potential 4-vector
$A_+^\mu$ is expressible in terms of the 4-velocity $u_+^\mu$. An
identical calculation can be performed to show that an analogous
result is true for negative charges.

\subsection{The Plane Electromagnetic Wave}

While in principle one can claim that all electromagnetic
configurations ultimately originate from the presence of charges,
there do exist nontrivial configurations in which no charges are
present, the most obvious and important example being that of the
electromagnetic wave. It is therefore important, both for this
reason and from a historical perspective, to show explicitly how
plane waves arise in the present context from the motion of the
relativistic continuum. We turn to this problem now.

Let us consider a plane electromagnetic wave with wave-vector
$\kappa$ travelling in the $x$-direction with the ${\bf E}$-field
plane-polarised in the $y$-direction. The 4-potential describing
this plane wave is,

\begin{equation}
A^\mu = (0, {\bf A}) = (0, 0, A_y\cos(\omega t-\kappa x), 0
)\,,\label{eqn:planewave}
\end{equation}
(where $\omega=c\kappa$), with corresponding ${\bf E}$ and ${\bf
B}$ fields,
\begin{equation}
{\bf E} = (0, E_y, 0) = (0, \kappa A_y\sin(\omega t-\kappa x),
0)\,,\qquad{\bf B} = (0, 0, B_z) = (0, 0, \kappa A_y\sin(\omega
t-kx))\,.\label{planeeb}
\end{equation}
We therefore seek solutions of the form,
\begin{equation}
A_+^\mu = (kc\gamma_+, k\gamma_+{\bf v}_+)\,,\qquad A_-^\mu =
(-kc\gamma_-, -k\gamma_-{\bf v}_-)\,.\label{eqn:pm}
\end{equation}
Applying (\ref{eqn:fourpot}) and equating with
(\ref{eqn:planewave}) we obtain the two conditions,
\begin{equation}
\gamma_+=\gamma_-\,,\qquad k(\gamma_+{\bf v}_+-\gamma_-{\bf v}_-)
= (0, A_y\cos(\omega t-\kappa x), 0) \,.\label{eqn:veccond}
\end{equation}

Ignoring equal velocity motions of the positive and negative
continua which have already been shown to have no electromagnetic
consequences, these conditions allow us to restrict our attention
to solutions of the form,

\begin{equation}
{\bf v}_+ = -{\bf v}_- = (0, v, 0)\,,\qquad\hbox{where}\quad
{v\over c} = {A\over\sqrt{A^2+4c^2}}\,,\label{eqn:soln}
\end{equation}
and we have defined $A=A_y\cos(\omega t-\kappa x)$ for
convenience. The velocities of the positive continuum and the
negative continuum here are equal in magnitude and opposite in
direction, so that there is no net charge, with the motion of both
being parallel to the electric field but $\mp\pi/2$ radians out of
phase respectively. It also follows from (\ref{eqn:soln}) that the
velocity of the continuum can never exceed the speed of light,
irrespective of the intensity of the plane wave. Substituting
(\ref{eqn:soln}) into (\ref{eqn:pm}) the motion of the continuum
is given by,

\begin{equation}
u_+^\mu = (\sqrt{c^2/k^2+A^2/4k^2}, 0, A/2k, 0)\,,\qquad u_-^\mu =
(\sqrt{c^2/k^2+A^2/4k^2}, 0, -A/2k, 0)\,.\label{eqn:pmsoln}
\end{equation}

These equations clearly show that the propagation of a plane
electromagnetic wave is described by the oscillation of the medium
in the direction of the electric field - the positive continuum
oscillates $\pi/2$ out of phase with {\bf E} while the negative
continuum oscillates with the same magnitude and precisely the
opposite phase. Thus the propagation of electromagnetic waves is
seen to be a direct manifestation of the oscillations of the
underlying relativistic continuum.

\subsection{Gauge Redundancies and the Principle of Superposition}

While the usual principle of superposition obviously still holds
for the 4-potential, we can now supplement this with the following
continuum-gauge-inspired superposition principle.

Consider two 4-potential fields $A^\mu = (c\gamma_+-c\gamma_-,
\gamma_+{\bf v}_+-\gamma_-{\bf v}_-)$ and $A'^\mu =
(c\gamma_+'-c\gamma_-', \gamma_+'{\bf v}_+'-\gamma_-'{\bf v}_-')$
in the continuum gauge which describe two different 4-velocity
field configurations. Then the superposition of the two field
configurations is described by the 4-potential $A''^\mu =
(c\gamma_+''-c\gamma_-'', \gamma_+''{\bf v}_+''-\gamma_-''{\bf
v}_-'')$ where the velocity vector field ${\bf v}_+''$
(respectively ${\bf v}_-''$) is given by the pointwise
relativistic sum of ${\bf v}_+$ and ${\bf v}_+'$ (respectively
${\bf v}_-$ and ${\bf v}_-'$),

\begin{equation}
{\bf v}_\pm'' = {{\bf v}_\pm + {\bf v}_\pm'\over1+{\bf
v}_\pm\cdot{\bf v}_\pm'/c^2}\,.\label{eqn:superpose}
\end{equation}

As mentioned earlier, the description of an electromagnetic
configuration in terms of 4-velocities $u_+^\mu$ and $u_-^\mu$ is
far from unique, as for each of the infinite number of
4-potentials $A^\mu=(\phi,{\bf A})$ with $|\phi|\leq|{\bf A}|$
describing that particular configuration, there exists an entire
two-parameter set of solutions at each point.

Recall the particular gauge choice in which $\phi=0$ everywhere.
We saw that the simplest `lowest energy' solution is given in this
case by ${\bf v}_+ = - {\bf v}_- = {\bf A}/2k$. However, we also
saw that it is possible to add, relativistically in the sense of
(\ref{eqn:superpose}), the same, arbitrary, possibly
time-dependent, 3-velocity vector field to both ${\bf v}_+$ and
${\bf v}_-$ without changing the 4-potential. If these velocity
fields have a real physical meaning then this additional freedom
will correspond to a large class of vacuum configurations which
can perhaps be interpreted in terms of the motion of an
arbitrarily distributed `Dirac sea' of particles and
antiparticles. This provides a means of adding energy density to
the vacuum without any observable electromagnetic effects.

\section{The Continuum as a Relativistic Fluid}

In this section we show that the spacetime continuum must be a
relativistic fluid of massive discrete particles, and that
interactions between sources and sinks give rise to the Lorentz
force equation. The fact that both Maxwell's equations and the
Lorentz force are consequences of the relativistic fluid model is
a strong indication that there is more to this description than
mere formalism, and that classical electrodynamics may in reality
have a fluid dynamical basis.

\subsection{The Massive Continuum}

We saw in (\ref{eqn:vcfinal}) that the velocity of the continuum
decreases with radius outside of the point charge acting as its
source. Had the continuum been massless, its velocity would have
been constant and equal to $c$ everywhere. We therefore conclude
that the continuum has mass and that there is an attractive
central force acting on the continuum outside of the charge.

It is possible to derive an expression for this attractive central
force. In particular, if we assume the charged particle is centred
at the origin, then the force ${\bf f}^i$ acting on an
infinitesimal element of the continuum at radius $r$ must
satisfy\cite{Landau75},

\begin{equation}
{\bf f}^i = {d{\bf p}^i\over dt} = m\gamma^3{d{\bf v}^i\over
dt}\,,\label{eqn:relforce}
\end{equation}
where $m=\rho_m \delta V$ is the mass of the test element assuming
that it has mass density $\rho_m$ and occupies volume $\delta V$.
To find the value of $dv/dt$, solve (\ref{eqn:vcfinal}) for and
differentiate the resulting equation with respect to $t$ to find
an expression for $dv/dt$ in terms of $v$. Rearranging terms and
simplifying, the field at radius $r$ is found to have the form,

\begin{equation}
\gamma^3{dv\over dt} = -{qc\over kr^2}\,.\label{eqn:gravfield}
\end{equation}

Thus there appears to be a Coulombic attraction between the charge
and the continuum around it, with the continuum having a
charge-to-mass ratio of $-c/k$. This is quite mysterious as in our
model charge is defined in terms of the motion of the continuum,
so clearly the continuum itself \emph{cannot} be charged. The
mystery will be resolved in due course.

Assuming continuum conservation, the continuum density $\rho_n$
will satisfy the following continuity equation,

\begin{equation}
\partial_\mu(\rho_nv^\mu) = {\partial(\rho_n\gamma)\over\partial t}
- \nabla\cdot(\rho_n\gamma{\bf v}) = 0\quad\Rightarrow\quad{1\over
r^2}{\partial\over\partial r}(r^2 \rho_n \gamma v)=0
\,.\label{eqn:radcont}
\end{equation}
where we ignore the time-derivative term as the system is in a
steady state condition, and use the rotational symmetry to rewrite
the divergence term in its spherical polar form. The solution is,

\begin{equation}
\rho_n = {S\over4\pi r^2\gamma v}\,,\label{eqn:contdenst}
\end{equation}
where $S$ is a radius-independent proportionality factor. Now, the
flux of continuum passing through a spherical shell at radius $r$ is
just $\Phi = 4\pi r^2\rho_n\gamma v$ (where the factor of $\gamma$
takes into account to the relativistic contraction in the radial
direction). But this is precisely the constant $S$ in
\eqref{eqn:contdenst} which can therefore be identified as the
strength of the charged particle sink/source.

\subsection{The Relativistic Fluid}

We discovered in the previous subsection that there is an
inverse-square law attraction of elements of the continuum towards
the point charge. Given that the continuum density is greater
closer to the charged particle, let us investigate the possibility
that the continuum may be a continuous, compressible, medium whose
attractive self-interactions result in the observed attraction. If
the attractive force between two volume elements of the continuum
is given by,

\begin{equation}
dF({\bf r}_1,{\bf r}_2) \sim \rho_n({\bf r}_1) \rho_n({\bf r}_2)
f(|{\bf r}_1-{\bf r}_2|)\,.\label{eqn:selfint}
\end{equation}
where $f(r)$ is some polynomial in $r$, then a little calculation
shows that an inverse square attraction is possible only if
$f(r)\sim r^{-4}$. However, the magnitude of the resultant force
on any element turns out to be infinitely large in this case.

There are three sources of these (logarithmic) divergences - (i)
the contribution from the core of the point charge where the
continuum density becomes infinite, (ii) the contribution from the
continuum at infinity, and (iii) the contribution from continuum
elements in the immediate neighbourhood of that element. The first
of these can be avoided if the charges are not pointlike, the
second can be avoided if the universe is either bounded or
homogeneous, and the third can be avoided by discarding the idea
that the continuum is some kind of continuous elastic medium, but
rather consists of a fluid of interacting discrete particles.

We are therefore led to conclude that our relativistic continuum
is a space-filling relativistic fluid and the electromagnetic
4-potential must be defined in terms of the ensemble motion of the
fluid as opposed to the motion of the individual discrete
particles. If the instantaneous fluid velocity ar $x^\mu$ is
$\zeta^\mu(x)$, then the 4-velocity appearing in
\eqref{eqn:fourpot} is,

\begin{equation}
u^\mu(x) = <\zeta^\mu>\,,\label{eqn:ensemble}
\end{equation}
where $<\zeta^\mu>$ indicates the time-averaged motion of the
particles in the neighbourhood of $x^\mu$. All other
electrodynamic quantities must be defined as time-averages in the
same way.

Although the configuration representing a charged particle is in
steady state, the fluid itself remains in constant motion. Recall
that the motion of an individual particle in the co-moving frame
is described by the total derivative\cite{Liboff98},

\begin{equation}
{d\zeta^\mu\over d\tau} = (\zeta^\nu\partial_\nu) \zeta^\mu =
(\zeta_\nu\partial^\nu) \zeta^\mu - {1\over2}
\partial^\mu(\zeta^\nu \zeta_\nu) = - (\partial^\mu \zeta^\nu -
\partial^\nu \zeta^\mu) \zeta_\nu\,,\label{eqn:comoving}
\end{equation}
where we have added a vanishing term using the fact that
$\zeta_\mu \zeta^\mu=c^2$. If we now consider the time-averaged
version of \eqref{eqn:comoving} and recall the definitions
\eqref{eqn:ensemble}, \eqref{eqn:fourpot} and
\eqref{eqn:stresstensor}, we find that,

\begin{equation}
k{du^\mu\over d\tau} = - F^{\mu\nu}
u_\nu\,,\label{eqn:lorentzfluid}
\end{equation}
which is in the form of the Lorentz force equation. In particular
we find that, on average, each particle moves \emph{as if} it were
charged with $q/m=-c/k$. This is precisely the charge-to-mass
ratio observed in the Coulomb-like attraction of
\eqref{eqn:gravfield}, and so the earlier mystery has been
resolved. Because \eqref{eqn:comoving} is a basic identity valid
for any motion of the relativistic fluid, this conclusion holds
irrespective of the precise nature of the interactions between the
fluid particles.

As further compelling evidence that classical electrodynamics has
relativistic fluid dynamics as its basis, we will now show that
the Lorentz force equation emerges automatically from the
interaction between sources and sinks when they are \emph{not}
assumed to be fixed in position.

The integral momentum equation for a fluid tells us that the force
on a target charged particle with charge $Q'$ due to a source
particle of charge $Q$ at distance $r$ is given by the rate of
change of momentum transfer to the target by the particles
entering or leaving the source. If we suppose that the target
particle has an effective radius $R$ then, assuming spherical
symmetry, it will have an effective volume of ${4\over3}\pi R^3$.
In accordance with (\ref{eqn:contdenst}), the density of fluid
particles encountering the target at distance $r$ from the source
is $\rho_n(r)$. If we further assume that each fluid particle is
identical with mass $m$, then the 3-momentum carried by each is
given by $m\gamma v$. Finally, the collision rate will be
determined by the strength $S'$ of the target. Thus the force on
the target will be given by the product of these contributions,

\begin{equation}
{\bf F} = {4\over3}\pi R^3 \rho_n m \gamma v S' = {mSS'R^3\over3
r^2} \,,\label{eqn:intmom}
\end{equation}
where he have used (\ref{eqn:contdenst}). This takes precisely the
form of Coulomb's law if we make the following identification,

\begin{equation}
Q = S\sqrt{mR^3\over3}\,,\label{eqn:chargedef}
\end{equation}
where the charge $Q$ is expressed in terms of the strength of the
source $S$, the mass $m$ of the fluid particles and the effective
charge radius $R$. Clearly for (\ref{eqn:intmom}) to hold,
positive charges must effectively act as sinks, and negative
charges as sources, or vice versa\footnote{The fact that equal
velocity contributions of fluid particles from the positive and
negative continua have no electromagnetic effects means that the
net momentum transfer must be zero, which in turn implies that
particles in the negative continuum must have equal and opposite
mass to those in the positive continuum. The constant $k$ in
\eqref{eqn:fourpot} must then be proportional to the mass of the
fluid particles, so that the 4-potential $A^\mu$ is nothing but
the net 4-momentum of the two-component fluid. This is a radically
different interpretation of the 4-potential from the one we are
accustomed to.}. The validity of Coulomb's law in turn implies the
validity of the Lorentz force equation\cite{Einstein20}, as we
have assumed from the outset that relativity holds. This completes
our description of classical electrodynamics.

\subsection{Stochastic Quantum Mechanics from the Self-Gravitating
Fluid}

We have shown that classical electrodynamics can be explained at
the macroscopic level in terms of the ensemble motion of a
relativistic fluid of massive (i.e. gravitationally interacting),
discrete particles. At a microscopic level this allows us to make
an explicit identification of our model with the stochastic
formulation of quantum mechanics.

In his monograph\cite{Nelson85}, Nelson gave a detailed derivation
of quantum mechanics on the basis of the conservative diffusion of
a classical fluid, wherein the Schr\"odinger wavefunction is
identified with the density of the fluid thus,

\begin{equation}
\psi = \sqrt{\rho} \, e^{iS/\hbar}\,,\label{eqn:wavefn}
\end{equation}
where $S$ is the stochastic analogue of Hamilton's principle
function. However there remained a number of important unresolved
problems. The first of these was \emph{``...to find a classical
Lagrangian, of system + background field oscillators +
interaction, that [...] produces a conservative diffusion
system."} Traditionally the system is assumed to be coupled in
some way to an electromagnetic background\cite{Pena95}. We have
succeeded here in showing that the classical fluid {\it is} the
electromagnetic background which Nelson sought.

Neither was he able to come to any definite conclusion about the
nature of the interparticle interactions responsible for the
conservative diffusion - except that it could {\it not} be
gravitational on dimensional grounds and that it was possibly of
electromagnetic origin. This second issue has also been resolved
here as gravitational interactions are themselves responsible for
electromagnetism. The possibility of a gravitational explanation
for the conservative diffusion, and consequently for the observed
magnitude of Planck's constant, has also been proposed by
Calogero\cite{Calogero97}.

\section{Summary and Conclusion}

We have demonstrated the simple yet profound result that all of
the equations of classical electrodynamics follow from the motion
of a two-component relativistic continuum satisfying the standard
equations of relativistic fluid dynamics. Charged particles appear
as sources and sinks of the continuum in this framework, while
electromagnetic waves are associated with oscillations of the
continuum. There is a freedom inherent in the 4-velocity
description of electrodynamics which can potentially account for
'dark matter'. The identification of the vacuum as a
self-gravitating fluid of discrete particles makes possible an
explicit connection to Nelson's stochastic formulation of quantum
mechanics, raising the tantalising prospect of having a natural,
unified, description of gravitation and quantum electrodynamics
purely in terms of classical general relativity.

Maxwell and others had struggled to find a mathematical
description of the underlying medium, the `aether', in which
electromagnetic waves were presumed to propagate. Although the
continuum we have described is not precisely equivalent to the
notion which the earlier proponents had had in mind, our analysis
does show that a description of electrodynamics in terms of an
underlying continuum is possible. This is particularly important
as the failure to find such a formulation historically contributed
to the origin of the concept of `fields' postulated not to require
such a medium. The field concept may not have been necessary after
all.

A number of issues still remain open. We are still left to ponder
the existence, interpretation and physical properties of the fluid
particles and their sources and sinks, to explain the origin of
quantised mass and charge, and the presence of the two continuum
components. Considerable evidence has already been gathered that
each of these issues can be resolved completely within the
framework of general relativity, and we feel that a unified
classical description of quantum theory and gravity is now close
at hand. It would certainly be fitting if Einstein's dream were
finally to be realised on the 100th anniversary of the birth of
his theory of relativity.


\begin{theacknowledgments}
I would like to thank Steve Carlip for his helpful comments on the
continuum gauge, to Mark Drela for his help with source-sink
interactions, and to Mark Hadley for clarifying an issue with time
nonorientability. I have benefitted from discussions with Andre
Gsponer, Abhas Mitra, Mahbub Majumdar and Asif Khalak at various
stages during the preparation of this work. I would also like to
acknowledge Eugen Negut, whose original ideas first aroused my
interest in this line of research. I would like to express my
gratitude to Chris Isham and Kellogg Stelle for graciously
allowing me to make use of the excellent facilities at Imperial
College where the first half of this work was completed, and to
Transport for London for the Jubilee line underground service
between Canons Park and Canary Wharf, with its many delays, where
most of the second half was conceived. Finally, thanks are due to
Roy Pike and to John David Jackson. None of these individuals are
responsible in any way for any errors which may remain.
\end{theacknowledgments}



\bibliographystyle{aipproc}   


%


\end{document}